# Big Data Technology Literature Review

Michael Bar-Sinai
Data Science Team
Institute for Quantitative Social Science at Harvard University
April 2013 (amended June 2015)

Storing and manipulating Big Data relies on various data structures, algorithms and technologies. Some of these are new, while others have existed for quite a while (the Bloom filter was presented in 1970) and are now making their way into mainstream software engineering. We present those algorithms and technologies briefly, and provide citations for further reading. This paper is organized as follows: First we discuss relevant theoretical background, including algorithms and design approaches. Then we look at concrete technologies and a few example products. Many Big Data products exist today, and new ones are being added constantly. Thus, products mentions below should be treated as concrete examples, rather than recommendations. Each topic ends with a "further reading" section, for those who wish to deepen their knowledge.

## Algorithms and Architectures

Traditionally, when discussing an algorithm, there is an underlying assumption that the data can fit in the memory. That is not to say that memory consumption is ignored - on the contrary, big-O space analysis is part of any introduction to algorithm class. But in the "big data" world, even algorithms that need $O(n)$ space have to be distributed.

### Sharding and Consistent Hashing

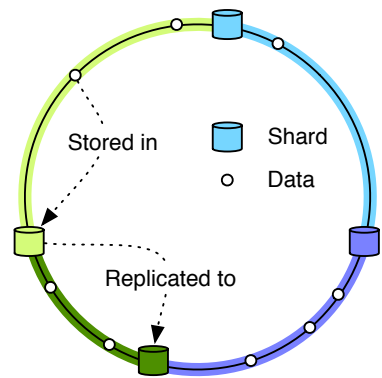

When the dataset is too large to be stored on a single machine, is has to be divided between a few machines. Such division is called horizontal partitioning, or "sharding". Each partial dataset stored on a single machine is called "a shard". Consistent Hashing is a good algorithm for sharding data. It is simple, scalable, and allows for easy fault tolerance. Consistent hashing is based on a hash function that maps each data point to a point on an edge of a circle. Shards are assigned points on the same circle. Each shard is then responsible for all data points in the arc immediately clockwise adjacent to it.

To allow fault tolerance, data are from shard *s* are backed up to shards further from the arc *s* is responsible for. This way, when *s* fails, the system can continue without losing data or dealing with a "corner case". Shards can be moved along the circle in order to balance their load.

*Further Reading*

**Bloom Filter**

The longest searches are those that fail. In a balanced binary tree, for example, a search for an absent key would have to look at an entire root-to-node path (O(log n)) in order to determine that the key is not present in the tree. A search for an existing key may be shorter. Bloom filter reduces some failed searches to O(1), at the expense of allowing some false positives.

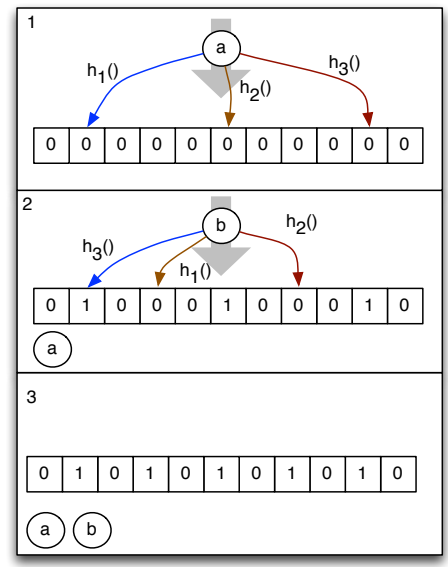

A Bloom filter consists of an array of bits, and a set of hash functions. When a key passes through the filter, its hash values are calculated according to each hash function. Then, the bits in the indices corresponding to the hash values are set to 1. In order to check whether a key may be in the filter, its hash values are calculated, and the bits in the corresponding indices are checked. If at least one bit is set to 0 - the key did not pass through the filter.

False positives, where a filter seems as if a certain key has passed through it while in fact it did not, are possible. False negatives, on the other hand, are impossible. The probability for false positives can be tuned by altering the amount of hash functions and the length of the bit array.

*Further Reading*

Bloom, B. H. (1970). Space/time trade-offs in hashing coding with allowable errors. *Communications of the ACM*, *13*(7), 422–426. doi:10.1145/362686.362692

**The Actor Model**

The Actor Model is a concurrent computation model where computation is done by actors that perform local calculations and send messages to each other. Actors do not share state in any way, thus it is possible to have many actors in a system without using locks, mutexes and other synchronization primitives.

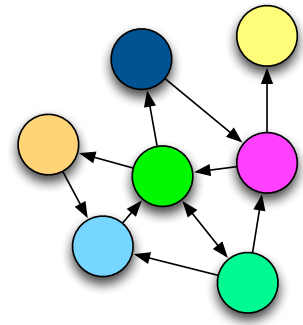

Actors are used today in many internet-scale systems, where numerous processes are needed in order to process large amounts of data. Prominent users of the actor framework are WhatsApp (using Erlang) and Twitter (using akka and scala).

*Further Reading*

Hewitt, C., Bishop, P., & Steiger, R. (1973). A universal modular ACTOR formalism for artificial intelligence, 235–245. Retrieved from http://dl.acm.org/citation.cfm?id=1624775.1624804

What'sApp team, *1 million is so 2011*, http://blog.whatsapp.com/?p=196
Waiming Mok, *How Twitter is Scaling*, https://waimingmok.wordpress.com/2009/06/27/how-twitter-is-scaling/

**Map-Reduce**

A way of implementing massively parallel computations, the Map-Reduce algorithm family breaks problems into two primitives: a Map function, mapping data to (key,value) pairs, and a Reduce function reducing a (key, list-of-values) pairs to (key, result) pairs. The values are aggregated in lists by the underlying Map-Reduce system. As all mappings and reductions are independent and stateless, they can be executed concurrently, possibly on different physical machines.

The "hello, world" example of map-reduce is counting appearances of words in a huge corpus (typically, the internet). In this case, the map function maps each word $w$ to ($w$, 1). The reduce function takes ($w$, [1,1,1,1,....1]) pairs and emits ($w$,$c$), where $c$ is the sum of the [1,1,1,...1] vector.

Some Map-Reduce variants allow partial reductions, to allow further concurrency.

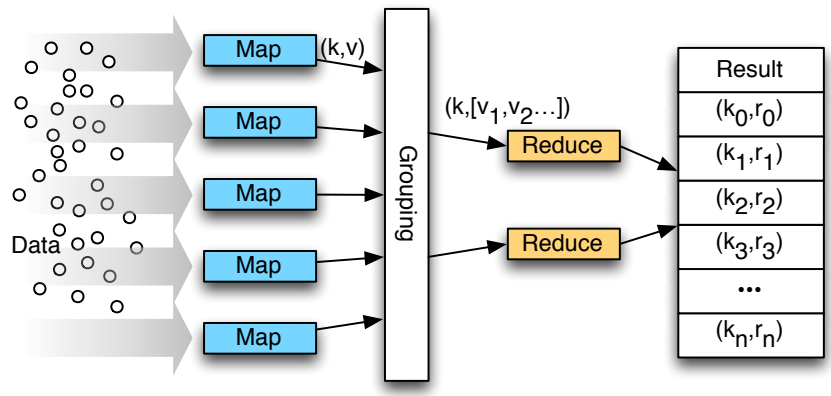

*Further Reading*

Dean, J., & Ghemawat, S. (2008). MapReduce: Simplified Data Processing on Large Clusters. (L. P. Daniel, Ed.)*Communications of the ACM*, *51*(1), 1–13. doi:10.1145/1327452.1327492

**Single Instruction on Multiple Data (SIMD)**
SIMD computations achieve parallelism by performing the same instruction on multiple data, simultaneously. This is made possible by hardware support. While most modern desktop CPUs have some SIMD ability, today this term is mostly used to refer to computations done by graphics processing units (GPU), which are designed to support this computation model.

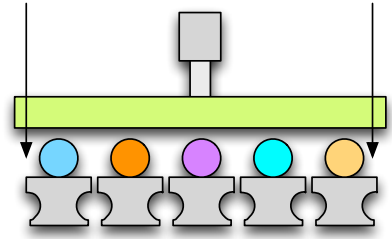

SIMD may involve low-level programming, but standards to allow higher-level programs are emerging, such as OpenCL. While not all algorithms may benefit from it, and while it is limited to a single machine, where SIMD can be applied it allows for extremely fast results execution times.

Our proposal includes a novel SIMD-based system called MapD. It uses CPU/GPU hybrid computations, splitting query execution between processing units in a way that leverages the strength of both. A paper describing it is in writing. A description of it is included in our proposal.

*Further Reading*
Jaskelainen, P. O., De La Lama, C. S., Huerta, P., & Takala, J. H. (2010). OpenCL-based design methodology for application-specific processors. *2010 International Conference on Embedded Computer Systems: Architectures, Modeling and Simulation* (pp. 223–230). IEEE. doi:10.1109/ICSAMOS.2010.5642061

Stokes, J. (2000). SIMD architectures. Retrieved March 28, 2013, from http://arstechnica.com/features/2000/03/simd/

**Sub-linear Algorithms**
Another approach to dealing with big data is to deal only with parts of it, while keeping an eye on the error bounds. Such algorithms may take time constraints or error bounds as part of their input. The sample itself can be calculated and maintained beforehand (such as in BlinkDB), or done as part of the calculation. Various sampling methods may be used, depending on
the hypothesis being estimated and the assumptions that could be made on the distributions in the dataset. Common sampling methods include random sampling (using true, hardware-based random generators), Gibbs sampling, and the Metropolis-Hastings algorithm.

*Further Reading*
Casella, G., & George, E. I. (1992). Explaining the Gibbs Sampler. *The American Statistician*, *46*(3), 167. doi:10.2307/2685208

Agarwal, S., Panda, A., Mozafari, B., Madden, S., & Stoica, I. (2012). BlinkDB: Queries with Bounded Errors and Bounded Response Times on Very Large Data. Databases; Distributed, Parallel, and Cluster Computing. Retrieved from http://arxiv.org/abs/1203.5485

HASTINGS, W. K. (1970). Monte Carlo sampling methods using Markov chains and their applications. *Biometrika*, *57*(1), 97–109. doi:10.1093/biomet/57.1.97

**REST**
Representational State Transfer (abbr. "REST" or "ReST") is an architectural style for communicating between software systems. It uses the HTTP protocol for communication, and thus offers a simple alternative to RPC, CORBA and SOAP/WSDL. Under REST, each data point has its own URL (universal resource locator, in a way restoring this term's original semantics). Client applications act on data points using their URLs and standard HTTP verbs, such as GET, POST and DELETE. Because of the ubiquity of HTTP support, systems that use "RESTful" architecture can be communicated from virtually any programming environment. First introduced by Roy Fielding in his doctoral thesis, REST is used today in many systems including Twitter, Flicker and Instagram in their public API. In the context of our proposal, using REST will allow us to abstract the storage layer and enable proxy servers and caches to speed up access times - as data points are rarely updated once published.

*Further Reading*
Roy Thomas Fielding. (2000). *Architectural Styles and the Design of Network-based Software Architectures*. University of California, Irvine. Retrieved from http://www.ics.uci.edu/~fielding/pubs/dissertation/top.htm
Dr. M. Elkstein, *Learn REST: A Tutorial*. http://rest.elkstein.org/2008/02/what-is-rest.html
Richardson, L., & Ruby, S. (n.d.). *RESTful Web Services*. Cambridge: O'Reilly Media.

**Storage Systems**
Big data cannot be stored on a single machine. Also, backup is an issue. Before we get into some of the ways it can be stored, it is important to note that this is a very active research area, and any system architecture should be able to cope with addition of new storage types. This is why we chose to build an additional interface on top of our storage layer.
Another issue to keep in mind is the CAP theorem. Sometimes called "Brewer's Theorem", it states that a storage implementation could choose any two properties from Consistency, Availability and Partition tolerance, but not all three. Big data storage systems can be divided according to which property they sacrifice in order to achieve the other two. Those that give up consistency are called "eventually consistent". In such systems, the data may not be consistent for a bounded amount of time. This is similar to the Internet's DNS system, where updates are propagated over 24 hours. Within the propagation period, different servers may return different replies to the same query.
Big-Dataverse Network is planned as a read-intensive system, where datasets are normally not deleted or altered (such actions are possible, but are not the main use case). Thus, the consistency property of the data store is of less importance - eventual consistency is sufficient.
Schema normalization often has to be sacrificed in order to achieve proper performance in big-data scenarios. Normalized schemas avoid data duplication and preserve model consistency, but rely on table joins at query time. Alas, joins cannot be used over big data, not only because of its size, but also since the data store is distributed. Thus, data modeling for big data allows some duplication in order achieve sufficient performance, at least for common queries.
Big data storage systems are grouped under the term "NoSQL", but vary in almost every field: data model, performance, API, and so on. Many NoSQL systems comparisons are available, we've included those we found to be more useful.

Some SQL databases are being adapted to support big data as well, pushing back against the NoSQL trend. But choosing a data store does not have to be an either/or decision; a dataset can be stored on multiple systems. This increasingly popular architectural style is called "polyglot persistence".

*Further Reading*

Cassandra vs MongoDB vs CouchDB vs Redis vs Riak vs HBase vs Couchbase vs Hypertable vs ElasticSearch vs Accumulo vs VoltDB vs Scalaris comparison. (n.d.). Retrieved March 28, 2013, from http://kkovacs.eu/cassandra-vs-mongodb-vs-couchdb-vs-redis

Cattell, R. (2011). Scalable SQL and NoSQL data stores. *ACM SIGMOD Record*, *39*(4), 12. doi:10.1145/1978915.1978919

Han, J. (2011). Survey on NoSQL database. *2011 6th International Conference on Pervasive Computing and Applications* (pp. 363–366). IEEE. doi:10.1109/ICPCA.2011.6106531

Brewer, E. A. (2000). Towards robust distributed systems (abstract). *Proceedings of the nineteenth annual ACM symposium on Principles of distributed computing - PODC '00* (p. 7). New York, New York, USA: ACM Press. doi:10.1145/343477.343502

Stonebraker, M. (2010). SQL databases v. NoSQL databases. *Communications of the ACM*, *53*(4), 10. doi:10.1145/1721654.1721659

**Key-Value Store: Project Voldemort**

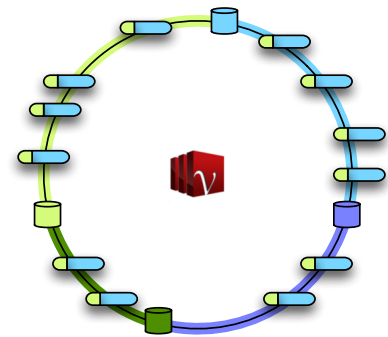

Key-Value stores are hash tables. Large, distributed, persistent, fault-tolerant hash tables. The keys,value pairs are distributed among the nodes using consistent hashing over the key. Replicated across nodes allow seamless fault tolerance. The data associated with each key is opaque, and thus cannot be queried - objects can only be retrieved by key.
While providing basic functionality, these data stores scale very well. Thus, when an application can infer key names from model data, they become very useful.
The original Key-Value store, named Dynamo, was developed by Amazon for various services, including their shopping carts. While Dynamo is proprietary, open-source versions have been developed. Project Voldemort is linkedIn's implementation, in use on their systems and others, including eBay and Mendeley.

DeCandia, G., Hastorun, D., Jampani, M., Kakulapati, G., Lakshman, A., Pilchin, A., Sivasubramanian, S., et al. (2007). Dynamo. *ACM SIGOPS Operating Systems Review*, *41*(6), 205. doi:10.1145/1323293.1294281

Sumbaly, R., Kreps, J., Gao, L., Feinberg, A., Soman, C., & Shah, S. (2012). Serving large-scale batch computed data with project voldemort. *FAST'12 Proceedings of the 10th USENIX conference on File and Storage Technologies*.

**Wide-Column Store: Cassandra**

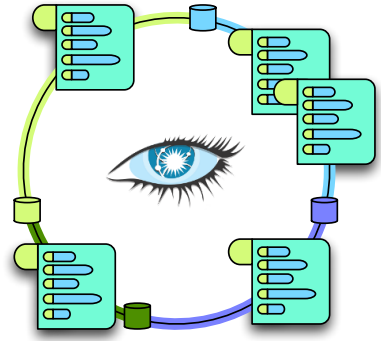

Cassandra is a column-based NoSQL database, able to store huge data sets. It has tunable performance and consistency, and has a relatively simple data model: each row consists of (key, value) pairs. Values may contain an additional level of (key, value) pairs, but these keys are not indexed. The number of columns for a single row can reach several millions (maximum is 2 billion), and may be dynamically changed - in fact, altering the columns of a row is part of common design patterns for data modeling using this data store.

Cassandra was developed by Facebook, and open sourced under the apache software foundation. It is based on Google's BigTable and Amazon's Dynamo, both proprietary systems. It uses consistent hashing to distribute the data. Each node has a bloom to reduce the time of failed searches. It is widely used today - http://planetcassandra.org/ boasts a long list of users, including netflix, reddit, twitter, symantec and adobe.

*Further Reading*

Lakshman, A., & Malik, P. (2010). Cassandra: a decentralized structured storage system. *ACM SIGOPS Operating Systems Review*. Retrieved from http://dl.acm.org/citation.cfm?id=1773922

Hewitt, E. (2011). *Cassandra: The Definitive Guide*. Cambridge: O'Reilly Media.

Chang, F., Dean, J., Ghemawat, S., Hsieh, W. C., Wallach, D. A., Burrows, M., Chandra, T., et al. (2008). Bigtable: A Distributed Storage System for Structured Data. *Sports Illustrated*, *26*(2), 1–26. doi:10.1145/1365815.1365816

**Relational DB Sharding**

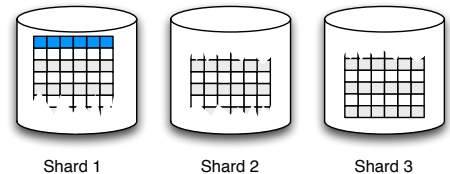

A "sharded" relational database has at least one of its tables split between different machines. This early form of big data storage is still in use today. While it uses relational databases, it sacrifices many of their fortes: built-in consistency, atomicity, joins, etc. This is the expected result of the CAP theorem. However, when a system does not need these properties, or could be designed to work around these limitations, relational DB sharding allows for big data storage using SQL databases, which are familiar and well-proven tools.

*Further Reading*

Rahul, R. (2008). *Shard - a Database Design*. http://technoroy.blogspot.com/2008/07/shard-database-design.html

Ries, E. (2009). *Sharding for Startups*. http://www.startuplessonslearned.com/2009/01/sharding-for-startups.html

**Document Store: MongoDB**
MongoDB is a NoSQL database which uses the "document model" to store data. In this model, loosely based around the JSON format, records consist of (key,value) pairs. Keys are strings, and values may be numbers, strings, arrays or objects. As objects are themselves sets of (key,value) pairs, this structure allows for unbounded nesting.
MongoDB has native support for map-reduce, text search and data sharding. It is in wide use today - Prominent users include the New York Times, eBay, intuit, github, and more.

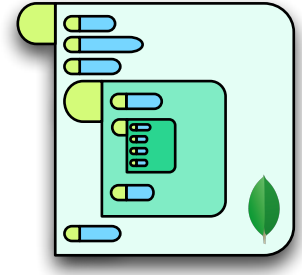

*Further Reading*
MongoDB site http://docs.mongodb.org/manual/
Mongo DB users list http://www.mongodb.org/about/production-deployments/
Bonnet, L., Laurent, A., Sala, M., Laurent, B., & Sicard, N. (2011). Reduce, You Say: What NoSQL Can Do for Data Aggregation and BI in Large Repositories. *2011 22nd International Workshop on Database and Expert Systems Applications* (pp. 483–488). IEEE. doi:10.1109/DEXA.2011.71

Banker, K. (2011). *MongoDB in Action*. Greenwich, CT, USA: Manning Publications Co.

**Graph Database: Neo4j**
Graph databases store graphs - nodes and directed edges, with properties for both. As graph theory is central to computer science, these data stores have many uses. Numerous libraries with implementations of graph algorithms are available - some databases come with common algorithms built in.
Neo4j is a popular graph database. It has its own query language, called "Cypher", and is fully open-source.

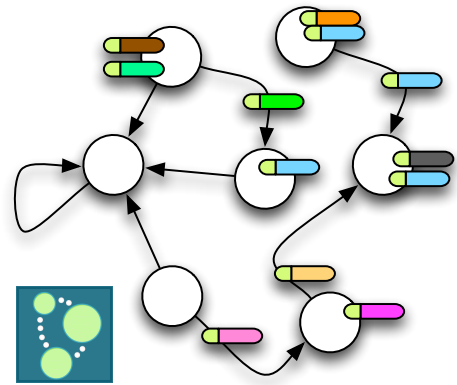

*Further Reading*
Angles, R., & Gutierrez, C. (2008). Survey of graph database models. *ACM Computing Surveys*, *40*(1), 1–39. doi:10.1145/1322432.1322433
Robinson, I., Webber, J., & Eifrem, E. (2013). *Graph Databases* (Early Release). O'Reilly Media.

Graph algorithms on Neo4J:
　　　http://api.neo4j.org/current/org/neo4j/graphalgo/GraphAlgoFactory.html
　　　http://docs.neo4j.org/chunked/snapshot/graph-algo.html

**Polyglot Persistence: All Together Now**
As each data store has its own set of merits, it sometimes makes sense to "mix and match" them. When the data is accessed using an API, clients do not interact directly with the data store. Thus, it is possible to split datasets between different types of data stores and account for the added complexity at the system servers, without affecting the clients at all.
For example, the Connectome project keep large images, and their meta-data. A possible combination would be storing the meta-data in a wide-column store, such as Cassandra, and storing the image data itself in a key-value store, such as Project Voldemort. This would preserve the functionality, but allow smaller rows in Cassandra, boosting its cache performance, and possibly improving the system performance, even though two systems have to be queried.
If, instead of Cassandra, we would have chosen MongoDB, such split would be a must; MongoDB documents cannot exceed 16MB.

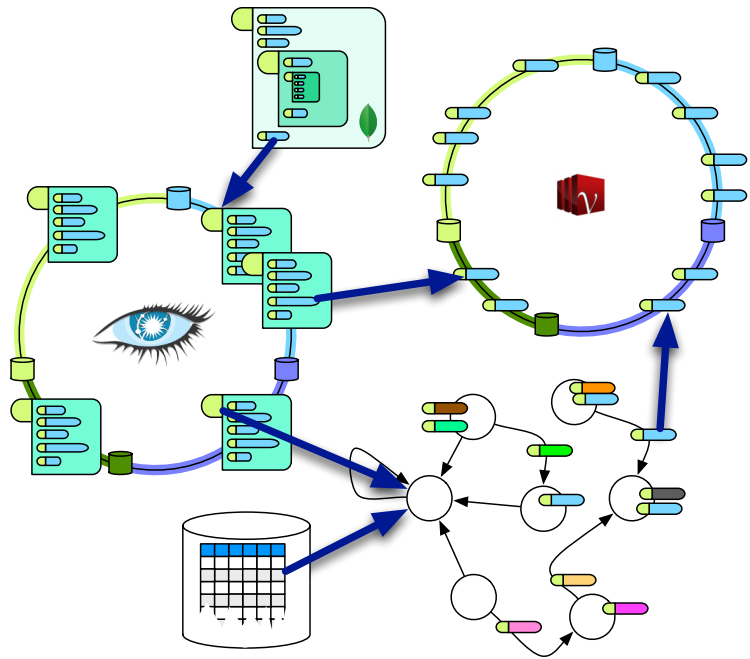

*Further Reading*
Fowler, M., & Sadalage, P. J. (2012). *Introduction to Polyglot Persistence: Using Different Data Storage Technologies for Varying Data Storage Needs* (p. 192). Boston, MA: Addison-Wesley Professional.

**Final Point: Privacy**
NoSQL/big data systems are often used for analyzing data harvested from social networks. It is imperative that we remember that often the "data subjects" of these datasets are people. Such dataset are usualy bound by data use agreements, guided by user consent. These legal instruments have to be respected, not only because breaching them might expose the researcher to lawsuits, but because betraying the trust of the data providers – the social networks and, ultimately, their human users – may mean it would be much harder for social scientists to be granted access to these valuable datasets. When in doubt about how to properly handle a dataset, consider consulting a lawyer familiar with the privacy legislation, an IRB, or the DataTags tool (www.datatags.org), developed by the Institute for Quantitative Social Science. DataTags is currently (June 2015) in beta, but already covers a few federal laws, including HIPAA, FERPA and CIPSEA.